\documentclass[showpacs,amsmath,amssymb,11pt]{article}
\usepackage{graphicx,color}
\usepackage{amsmath}
\usepackage{amssymb}
\usepackage{amsmath,amsfonts,latexsym,graphicx,amssymb}
\usepackage{amsfonts}
\usepackage{amssymb}
\usepackage{mathrsfs}
\usepackage{amsthm}
\usepackage[utf8]{inputenc}
\newcommand{\ot}{{\,\otimes\,}}
\usepackage{graphicx}
\usepackage{amsmath}
\newcommand{\Md}{M_d(\mathbb{C})}

\newtheorem{Remark}{Remark}

\newtheorem{Proposition}{Proposition}
\date{}
\def\oper{{\mathchoice{\rm 1\mskip-4mu l}{\rm 1\mskip-4mu l}%
{\rm 1\mskip-4.5mu l}{\rm 1\mskip-5mu l}}}
\def\<{\langle}
\def\>{\rangle}

\begin{document}
\title{\textbf{
A class of bistochastic positive optimal maps in $M_d(\mathbb{C})$}}
\author{Adam Rutkowski$^{1,2}$, Gniewomir Sarbicki$^3$, and Dariusz Chru\'sci\'nski$^3$
 \\
$^1$ Faculty of Mathematics, Physics and Informatics,\\ University of Gda{\'n}%
sk, 80-952 Gda{\'n}sk, Poland\\
$^2$ Quantum Information Centre of Gda\'{n}sk, 81-824 Sopot, Poland  \\
$^3$ Institute of Physics, Faculty of Physics, Astronomy and Informatics, \\
Nicolaus Copernicus University, Grudzi{a}dzka 5, 87--100 Toru\'n, Poland }

\maketitle

\begin{abstract}
We provide a straightforward generalization of a positive map in $M_3(\mathbb{C})$ considered recently by Miller and Olkiewicz \cite{Miller}.
It is proved that these maps are optimal and indecomposable. As a byproduct we provide a class of PPT entangled states in $d \ot d$.
\end{abstract}

Positive maps in matrix algebras play important role both in mathematics and theoretical physics \cite{I,Bhatia,III,TOPICAL}. In the recent paper \cite{Miller} paper Miller and Olkiewicz considered a  linear map $\Lambda_3 : M_3(\mathbb{C}) \rightarrow M_3(\mathbb{C})$ ($\Md$ denotes a matrix algebra of $d \times d$ complex matrices) defined as follows

\begin{equation}\label{Miller}
 \Lambda_{3} \left(\begin{array}{ccc}
a_{11} & a_{12} & a_{13}\\
a_{21} & a_{22} & a_{23}\\
a_{31} & a_{32} & a_{33}
\end{array}\right)=\left(\begin{array}{ccc}
\frac{1}{2}\left(a_{11}+a_{22}\right) & 0 & \frac{1}{\sqrt{2}}a_{13}\\
0 & \frac{1}{2}\left(a_{11}+a_{22}\right) & \frac{1}{\sqrt{2}}a_{32}\\
\frac{1}{\sqrt{2}}a_{31} & \frac{1}{\sqrt{2}}a_{23} & a_{33}
\end{array}\right)\geq0 \ .
\end{equation}
It was proved \cite{Miller} that $\Lambda_3$ is a bistochastic positive extremal (even exposed) non-decomposable map.
In this paper we provide the following generalization $\Lambda_d : \Md \to \Md$:

\begin{equation}
\Lambda_{d}\left(A\right)=\frac{1}{d-1}\left(\begin{array}{ccccc}
{\displaystyle \sum_{i=1}^{d-1}}a_{ii} & \cdots & 0 & 0 & \sqrt{d-1}a_{1d}\\
\vdots & \ddots & \vdots & \vdots & \vdots\\
\vdots & \cdots & {\displaystyle \sum_{i=1}^{d-1}}a_{ii} & 0 & \sqrt{d-1}a_{d-2,d}\\
0 & \cdots & 0 & {\displaystyle \sum_{i=1}^{d-1}}a_{ii} & \sqrt{d-1}a_{d,d-1}\\
\sqrt{d-1}a_{d1} & \cdots & \sqrt{d-1}a_{d,d-2} & \sqrt{d-1}a_{d-1,d} & \left(d-1\right)a_{dd}
\end{array}\right)\ ,
\end{equation}
where $A = [a_{ij}] \in \Md$.

\begin{Proposition} $\Lambda_d$ is a positive map.
\end{Proposition}

Proof: let $y=\left(\begin{array}{c}
\mathbf{x}\\
x_{d}
\end{array}\right)\in\mathbb{C}^{d}, \mathbf{x} \in \mathbb{C}^{d-1}$ and  $P_{i}= |i\>\<i|$  for $i=1,\ldots,d-1$. One has

\[
\Lambda_{d}\left(yy^{\dagger}\right)=\frac{1}{d-1}\left(\begin{array}{c|c}
\left\Vert x\right\Vert ^{2}\mathbb{I}_{d-1} & \sqrt{d-1}\left(x_{d} P_{1}\mathbf{x} +\sum_{i=2}^{d-1}\bar{x}_{d} P_{i}\bar{\mathbf{x}}\right)\\
\hline \sqrt{d-1}\left(x_{d} P_{1}\mathbf{x} + \sum_{i=2}^{d-1}\bar{x}_{d} P_{i}\bar{\mathbf{x}}\right)^{\dagger} & \left(d-1\right)\left|x_{d}\right|^{2}
\end{array}\right) \ .
\]
Now we use the well known result \cite{Bhatia}: a block matrix
\begin{equation*}\label{}
  \left( \begin{array}{c|c} A & B \\ \hline B^\dagger & C \end{array} \right) ,
\end{equation*}
with $C > 0$ is positive iff
\begin{equation}\label{}
  A \geq B C^{-1} B^\dagger \ .
\end{equation}
Hence, to prove that $\Lambda_{d}\left(yy^{\dagger}\right) \geq 0$ it is necessary and sufficient to show that
\[
\left|x_{d}\right|^{2}\left\Vert x\right\Vert ^{2}\mathbb{I}_{d-1} - \left(x_{d} P_{1} \mathbf{x} + \sum_{i=2}^{d-1}\bar{x}_{d} P_{i} \bar{\mathbf{x}}\right) \left(x_{d}P_{1}\mathbf{x} + \sum_{i=2}^{d-1}\bar{x}_{d} P_{i}\bar{\mathbf{x}}\right)^{\dagger}\geq0 .
\]
One has

\begin{eqnarray*}
&&
\left(x_{d}P_{1}\mathbf{x}+\sum_{i=1}^{d-1}\bar{x}_{d} P_{i}\bar{\mathbf{x}}\right) \left(x_{d} P_{1}\mathbf{x} + \sum_{i=1}^{d-1}\bar{x}_{d} P_{i}\bar{\mathbf{x}}\right)^{\dagger} \leq \left\Vert \left(x_{d}p_{1}x+\sum_{i=2}^{d-1}\bar{x}_{d}p_{i}\bar{x}\right)\right\Vert ^{2}\mathbb{I}_{d-1} \\
&& = \left|x_{d}\right|^{2}\left(\left\Vert P_{1}\mathbf{x}\right\Vert^{2}+\sum_{i=2}^{d-1}\left\Vert P_{i}\bar{\mathbf{x}}\right\Vert^{2}\right) \mathbb{I}_{d-1}  = \left|x_{d}\right|^{2}\left\Vert \mathbf{x}\right\Vert ^{2}\mathbb{I}_{d-1} ,
\end{eqnarray*}
which ends the proof. \hfill $\Box$

\begin{Remark} It is very easy to check that $\Lambda_d$ is unital and trace-preserving and hence it defines a positive bistochastic map.
\end{Remark}


\begin{Proposition}  $\Lambda_d$ is nondecomposable.
\end{Proposition}
Proof: to prove it we construct a PPT state $\rho_{\rm PPT}$ such that ${\rm Tr}(W_d \rho) < 0$, where $W_d = (\oper \ot \Lambda_d)P^+_d$ denotes the corresponding entanglement witness (\cite{T1}  and the recent review \cite{TOPICAL}). Let us define

\begin{equation*}
\rho=\left(\begin{array}{c|c|c|c|c}
\sqrt{d-2}e_{11}+e_{dd} & 0 & \cdots & 0 & -e_{1d}\\
\hline 0 & \sqrt{d-2}e_{22}+e_{dd} & \cdots & 0 & -e_{2d}\\
\hline \vdots & \vdots & \ddots & \vdots & \vdots\\
\hline 0 & 0 & \cdots & \sqrt{d-2}e_{d-1,d-1}+e_{dd} & -e_{d-1,d}^{T}\\
\hline -e_{d1} & -e_{d2} & \cdots & -e_{d,d-1}^{T} & \mathbb{I}-\left(1-\sqrt{d-2}\right)e_{dd}
\end{array}\right) \ ,
\end{equation*}
where $e_{ij} = |i\>\<j|$. Let us observe that $\rho\geq0$ iff the following $d-1\times d-1$ submatrix

\begin{equation}
\left(\begin{array}{ccccc}
\sqrt{d-2} & 0 & \cdots & 0 & -1\\
0 & \sqrt{d-2} & \cdots & 0 & -1\\
\vdots & \vdots & \ddots & \vdots & \vdots\\
0 & 0 & \cdots & \sqrt{d-2} & -1\\
-1 & -1 & \cdots & -1 & \sqrt{d-2}
\end{array}\right) \, \geq\, 0 \ ,
\end{equation}
which is the case due to the fact that its eigenvalues read: $\{\lambda_1=0,\lambda_2=\sqrt{d-2},\lambda_3= 2\sqrt{d-2}\}$,
where $\lambda_1,\lambda_3$ are simple and $\lambda_2$ has multiplicity $d-3$. Consider now the partial transposed
\begin{equation*}
\rho^{\Gamma}=\left(\begin{array}{c|c|c|c|c}
\sqrt{d-2}e_{11}+e_{dd} & 0 & \cdots & 0 & -e_{d1}\\
\hline 0 & \sqrt{d-2}e_{22}+e_{dd} & \cdots & 0 & -e_{d2}\\
\hline \vdots & \vdots & \ddots & \vdots & \vdots\\
\hline 0 & 0 & \cdots & \sqrt{d-2}e_{d-1,d-1}+e_{dd} & -e_{d,d-1}^{T}\\
\hline -e_{1d} & -e_{2d} & \cdots & -e_{d-1,d}^{T} & \mathbb{I}-\left(1-\sqrt{d-2}\right)e_{dd}
\end{array}\right)\ .
\end{equation*}
Its positivity follows from the simple observation that the following $2 \times 2$ submatrices

\begin{equation}
\left(\begin{array}{cc}
\sqrt{d-2} & -1\\
-1 & \sqrt{d-2}
\end{array}\right)\qquad\text{and}\qquad\left(\begin{array}{cc}
1 & -1\\
-1 & 1
\end{array}\right)
\end{equation}
are positive.  Now,
\[
\text{Tr}\left(W_d \rho \right)=2\left(d-1\right)\left(\sqrt{d-2}-\sqrt{d-1}\right)<0 \ ,
\]
which finally proves that $\Lambda_d$ is nondecomposable. \hfill $\Box$

Now we are ready to show that a map $\Lambda_d$  is optimal \cite{Lew}.

\begin{Proposition} $\Lambda_d$ is optimal.
\end{Proposition}
Proof: to prove optimality we use the following result from \cite{Lew}: if the entanglement witness $W = (\oper \ot \Lambda)P^+_d$ allows for a set of product vectors $\psi_k \ot \phi_k$ such that
\begin{equation}\label{}
  \< \psi_k \ot \phi_k|W|\psi_k \ot \phi_k\> = 0 \ ,
\end{equation}
then if $\psi_k \ot \phi_k$ span $\mathbb{C}^d \ot \mathbb{C}^d$ the map $\Lambda$ is optimal. Now, take arbitrary $x \in \mathbb{C}^d$ and define
\begin{equation}\label{}
  W_d(x) = {\rm Tr}_1 (W_d \cdot  |x\>\<x| \ot \mathbb{I}_d ) .
\end{equation}
One finds
 \begin{equation}
   W_d(x) = \left[ \begin{array}{c|c} zI_{d-1} & \vec{a} \\ \hline \vec{a}^\dagger & u \end{array} \right],
 \end{equation}
 where
 \begin{displaymath}
  a_i = \sqrt{d-1} \cdot \left\{ \begin{array}{lll} x_d^* x_i  & \text{for} & i<d-1 \\ x_d x_i^*  & \text{for} & i=d-1 \end{array} \right.\ ,
 \end{displaymath}
 $z=\sum_{i=1}^{d-1} |x_i|^2$ and $u=(d-1)|x_d|^2$. Note that $W_d(x)$ is at least of rank $d-1$ and hence its kernel is at most 1-dimensional. To find the corresponding zero-mode of $W_d(x)$ we consider
 \begin{displaymath}
   {\rm det}\, W_d(x) = -(d-1) |x_d|^2 \left( \sum_{i=1}^{d-1} |x_i|^2 \right) \cdot z^{d-2} + (d-1) |x_d|^2 z^{d-1} = 0\ .
 \end{displaymath}
Observing that the last row of $W_d(x)$ is a combination of the previous ones, we find the vector of the kernel solving the equation
 \begin{equation}
  [zI | \vec{a}] \left[ \begin{array}{l} \vec{v} \\ w \end{array} \right] = z\vec{v} + \vec{a}w = 0
 \end{equation}
 which implies (up to a scalar), that $\vec{v}=\vec{a}$ and $w=-z$. Denoting the solution as $y(x)$, one gets the family $q(x) = x \otimes y(x)$ of product vectors such that $\<x \ot y(x)|W|x\ot y(x)\>=0$. A vector from the family has the following coordinates:
 \begin{displaymath}
  \begin{array}{ccccc}
    x_1 x_1 x_d^* & \dots & x_1 x_{d-2} x_d^* & x_1 x_{d-1}^* x_d & x_1 \sum_{i=1}^{d-1} x_i x_i^* \\
    x_2 x_1 x_d^* & \dots & x_2 x_{d-2} x_d^* & x_2 x_{d-1}^* x_d & x_2 \sum_{i=1}^{d-1} x_i x_i^* \\
    \vdots & \vdots & & \vdots \\
    x_d x_1 x_d^* & \dots & x_d x_{d-2} x_d^* & x_d x_{d-1}^* x_d & x_d \sum_{i=1}^{d-1} x_i x_i^* \ .
  \end{array}
\end{displaymath}
It remains to show that vectors $q(x)= x \ot y(x)$ span $\mathbb{C}^d \ot \mathbb{C}^d$. Suppose that there exists a vector $\alpha= \sum_{i,j=1^d} \alpha_{i,j} |e_i\> \ot |e_j\>$ orthogonal to $q(x)$ for all $x$, that is,
 \begin{displaymath}
  \sum_{i=1}^d \left( \sum_{j=1}^{d-2} \alpha_{i,j}^* x_i x_j x_d^* + \alpha_{i,d-1}^* x_i x_{d-1}^* x_d + \alpha_{i,d}^* x_i \left( \sum_{i=1}^{d-1} x_i x_i^* \right) \right) = 0\ .
 \end{displaymath}
We stress that in the linear space of polynomials of $2d$ variables $x_i$ and $x_i^*$ are linearly independent. The monomial $x_i x_1 x_1^*$ appears in the sum only once multiplied by the coefficient $\alpha_{i,d}$. Hence because different monomials are linearly independent in the space of polynomials one concludes that $\alpha_{i,d}=0$.  Next observe, that the monomial $x_i x_{d-1}^* x_d$ appears only once multiplied by the coefficient $\alpha_{i,d-1}$. Thus one concludes that $\alpha_{i,d-1}=0$.  Finally, we have to prove, that the sum $\sum_{i=1}^d \sum_{j=1}^{d-2} \alpha_{i,j}^* x_i x_j x_d^*$ is zero iff all coefficients are zero. Indeed, all the coefficients multiply the different monomials. There are no non-zero vectors orthogonal to the subspace spanned by the vectors $q(x)$, so these vectors span the whole Hilbert space of the system, what implies optimality of the witness. \hfill $\Box$


In conclusion we have shown how to generalize  a positive map in $M_3(\mathbb{C})$ considered  in \cite{Miller} to a positive map in  $M_d(\mathbb{C})$.
We have proved that this map is optimal and indecomposable. As a byproduct we provide a class of PPT entangled states in $d \ot d$. It would be interesting check whether this generalized map is extremal or even exposed.

\section*{Acknowledgements}
A. Rutkowski was supported by a postdoc internship decision number DEC\textendash{}2012/04/S/ST2/00002, from the Polish National Science Center.
D. Chru\'sci\'nski and G. Sarbicki were partially supported by the National Science Center project
DEC-2011/03/B/ST2/00136.

\end{document}